\documentstyle[preprint,aps,epsf]{revtex}
\newcommand{\beq}{\begin{equation}}
\newcommand{\eeq}{\end{equation}}
\newcommand{\ds}{\displaystyle}
\begin{document}
\tightenlines
\draft

\title{
 Local equilibrium in heavy-ion collisions. \\
 Microscopic analysis of a central cell versus infinite matter.
}
\author{
L.~V.~Bravina,$^{1,2}$ E.~E.~Zabrodin,$^{1,2}$ 
S.~A.~Bass,$^{3}$ M.~Bleicher,$^{4}$ M.~Brandstetter,$^{5}$ 
S.~Soff,$^{5}$ H.~St{\"o}cker,$^{5}$ and W.~Greiner$^{5}$
}
\address{
$^1$ Institut f\"ur Theoretische Physik, Universit\"at T\"ubingen,
Auf der Morgenstelle 14, D-72076 T\"ubingen, Germany \\
$^2$ Institute for Nuclear Physics, Moscow State University,
RU-119899 Moscow, Russia \\
$^3$ National Superconducting Cyclotron Laboratory, Michigan State
University, East Lansing, Michigan 48823 \\
$^4$ Nuclear Science Division, Lawrence Berkeley Laboratory, Berkeley,
California 94720 \\
$^5$ Institut f\"ur Theoretische Physik, Universit\"at Frankfurt,
Robert-Mayer-Str. 8-10, D-60325 Frankfurt, Germany \\ 
}

\maketitle

\begin{abstract}
We study the local equilibrium in the central $V = 125$ fm$^3$ cell 
in heavy-ion collisions at energies from 10.7{\it A} GeV (AGS) to 
160{\it A} GeV (SPS) calculated in the microscopic transport
model. In the present paper the hadron yields and energy spectra in 
the cell are compared with those of infinite nuclear matter, as
calculated within the same model. The agreement between the spectra
in the two systems is established for times $t \geq 10$ fm/$c$ in the 
central cell. The cell results do not deviate noticeably from the
infinite matter calculations with rising incident energy, 
in contrast to the apparent discrepancy with predictions 
of the statistical model (SM) of an ideal hadron gas.
The entropy of this state is found to be very close to the maximum 
entropy, while hadron abundances and energy spectra differ 
significantly from those of the SM. 
\end{abstract}


\widetext

\section{Introduction} 
\label{sec1}

The hypothesis that local equilibrium (LE) can occur in the 
system of two heavy nuclei colliding head on at relativistic energies
is one of the most intriguing problems in high-energy physics 
\cite{Fer50,Land53,SBM,Shur80,Bjor83,LE84,StGr86,GeKa93,Geig95,Cass90,
BrSt95,Bass98,Blei98,Bec98,UH99,LeRa99,ClRe99,SHSX99}. 
In this paper we continue to study the equilibration of hot and dense
hadronic matter, produced in the central (with volume $V = 125$ 
fm$^3$) cell in relativistic heavy-ion collisions, started in Ref. 
\cite{lep1} (see also Refs. \cite{lv98plb,lv99jpg}). The equilibration 
of infinite nuclear matter has been studied separately in Refs.
\cite{Belk98,Bran99} and also in Ref. \cite{Brat00}. Here we will 
compare this idealized scenario to the properties of matter in the 
central cell of a heavy-ion collision. Both, the dynamics of the 
nuclear collisions in the energy range from from 10.7 to 
160{\it A} GeV, as well as the infinite matter are calculated in the 
framework of the ultrarelativistic quantum molecular dynamics (UrQMD) 
model \cite{urqmd1,urqmd2}. Note that in Ref. \cite{Brat00} hadron 
string dynamics (HSD) model \cite{hsd} has been used for the infinite 
nuclear matter calculations.

The concept of local equilibrium assumes that for a given system a
state is reached where all deviations of the system characteristics 
from the equilibrium ones can be treated as small perturbations. For
instance, the distribution function of particles in the phase 
space $\{ \vec{x}, \vec{p} \}$, $f(\vec{x}, \vec{p})$, can be
approximated by 
\beq
\ds
f (\vec{x}, \vec{p}) = f^{\rm eq} (p)\, \left[1 + 
\Delta (\vec{x}, \vec{p})\right]\quad,
\label{eq1}
\eeq
where $\Delta (\vec{x}, \vec{p}) \ll 1$ and $f^{\rm eq}(p)$ is the
equilibrium distribution function. But the central cell in a
nuclear collision is an open system. Neither the energy density nor
the particle composition in the cell is conserved. To decide whether
or not the LE is attained in this open system the following procedure
has been developed \cite{lep1,lv98plb,lv99jpg}. At the early stage of
the collision the conditions of the {\it kinetic\/} equilibrium 
should be fulfilled. This implies the absence of significant (in the 
sense of LE definition) collective flow in the cell together with the 
nearly Maxwellian [see Eq.~(\ref{eq1})] shape of particle momentum
distributions. These conditions should be completed by the requirement
of {\it thermal\/} and {\it chemical\/} equilibrium. It means that
the particle energy spectra, $d N / 4 \pi p E d E$, and particle 
abundances should be close to those of equilibrated matter.
In Refs. \cite{lep1,lv98plb,lv99jpg} the statistical model (SM) of an 
ideal hadron gas \cite{Belk98} with essentially the same hadron 
species as in the UrQMD model has been applied to calculate the 
characteristics of hadron-resonance matter in equilibrium. As an input 
the SM needs only three parameters, extracted from the microscopic 
calculations in the cell. Namely, the energy density $\varepsilon$, 
the baryon density $\rho_{\rm B}$, and the strangeness density 
$\rho_{\rm S}$, have to be determined. Solving then a system of 
nonlinear equations \cite{lep1} one can reproduce particle yields and 
energy spectra in the SM via the temperature $T$, baryochemical 
potential $\mu_{\rm B}$, and strangeness chemical potential 
$\mu_{\rm S}$. If the spectra of hadrons in the microscopic cell 
calculations are close to the spectra predicted by the SM, the LE is 
assumed to set in in the cell.

The striking result of the investigations made in Ref. \cite{lep1} 
is that the hot (and initially dense) hadron matter in the central 
cell does not reach the pure equilibrium in the sense of statistical 
mechanics during the time evolution, i.e., the state of maximum
entropy \cite{LaLi80}. However, it reaches the stage of kinetic 
equilibrium at time $t \cong 10$ fm/$c$, almost irrespective of 
the energy of colliding nuclei. The deviations of the UrQMD results 
from the SM predictions become significant with rising 
center-of-mass energy of the system.

To make a final decision about the possibility of local equilibration
in a heavy-ion collision, simulated within a microscopic model,
the comparison with the equilibrated infinite matter, simulated 
within the same model, should be performed. This is essentially the 
subject of the present paper, which is organised as follows. A
brief description of the model applied and the results on relaxation 
of the infinite hadron matter are given in Sec.~\ref{sec2}. After a
certain nonstationary stage, yields and apparent temperatures of 
hadron species in the box with periodic boundary conditions, which
models infinite matter, become time independent. This means that
an equilibrium (or rather stationary) state is attained. 
Section \ref{sec3} presents the comparison of the dynamical cell 
calculations with the stationary phase of infinite hadron matter. It 
is shown that the spectra in the cell at $t \geq 10$ fm/$c$ agree 
well with the infinite matter calculations, in contrast to the SM 
predictions. The discussion of the results obtained is given in 
Sec.~\ref{sec4}. The origin of discrepancies between the UrQMD 
infinite matter simulations and the SM predictions is traced 
\cite{lv99jpg,Belk98} to many-body ($N \geq 3$) decays of resonances 
and processes of multiparticle production via the excitation and 
fragmentation of quark-diquark (or quark-antiquark) strings. 
In the present paper we argue that the absorption of the excess of 
particle production (mainly pions) by the resonances, which can
interact both inelastically and elastically within their lifetimes,
is the key process which maintains the total balance of hadrons in
the model calculations. The conclusions are drawn in Sec.~\ref{sec5}.

\section{Simulation of infinite hadron matter}
\label{sec2}

\subsection{Initial conditions}
\label{sub2a}

To simulate the infinite hadron gas in the UrQMD model a cubic box
of a volume $V = 5 $ fm $\times 5 $ fm $\times 5$ fm $ = 125$ fm$^3$ 
with periodic boundary conditions has been chosen 
\cite{Belk98,Bran99,Brat00,urqmd1}. These
conditions ensure that particles always remain in the box. For
instance, if one particle crosses the face of the box and leaves it,
another particle identical to the first one enters immediately the
box from the opposite box face. The box plays a role of a heat bath
in which the total energy density is conserved.

To define the initial state the number of particles, their
composition and momenta must be specified. It can be a baryon-free
gas of mesons, a baryon-antibaryon gas, or a gas of strings and
resonances. For infinite nuclear matter with zero net 
strangeness and nonzero net baryon charge it is convenient to
initialise a system consisting of protons and neutrons only
\cite{Bran99}. Here the nucleons are uniformly distributed in the
box volume. Their momenta are initialized randomly in a Fermi sphere
with the subsequent rescaling to provide the required energy
density. Further details concerning the simulations of infinite
nuclear matter can be found elsewhere \cite{Belk98,Bran99,Brat00}.

It is worth noting that such an analysis should not be mixed up with
the study of dynamics of equilibration in relativistic heavy-ion 
collisions. In the latter case two Lorentz-contracted nuclei pass 
through each other and produce a nearly cylindrically expanding volume 
of hot and dense quark-hadron matter. The equilibration process in the 
central cell of such system proceeds faster compared to the box with 
the same values of energy density, baryon density and strangeness 
density. The cell is an open system and the most energetic particles 
leave it freely, causing an effective cooling of the rest of the 
hadron matter in the cell. Times needed to reach thermal and chemical 
equilibration in the box (isolated system) are, apparently, much 
larger because of the permanent production of new particles in very 
energetic collisions at the nonequilibrium stage.

\subsection{Equilibration in the box}
\label{sub2b}

Criteria of equilibrium of hot hadronic matter in the box 
\cite{Belk98,Bran99} are similar to the criteria of equilibrium in 
the central cell of ultrarelativistic heavy-ion collisions, formulated 
in Refs. \cite{lep1,lv98plb,lv99jpg}. We recall them briefly. 
Conditions of {\it kinetic\/} equilibrium imply that (i) the 
velocity distributions $f_j(v)$ of hadron species $j$ are isotropic 
and obey Maxwell distribution
\beq
\ds
f_j (v) \propto \exp{\left( - \frac{m_j v^2}{2 T} \right)}
\label{eq2}
\eeq
with $T$ being the temperature of the system. For local equilibrium
some deviations [see Eq.~(\ref{eq1})] from the Maxwellian shape are
possible. Therefore, it is convenient (especially for the central cell
in heavy-ion collisions) to complete the verification procedure by
(ii) the requirement of isotropy in the pressure sector. As a matter 
of fact, full isotropy of the velocity distributions leads obviously 
to the isotropy of the diagonal elements of the pressure tensor. Thus, 
criteria (i) and (ii) are not independent. But, instead of checking 
isotropy of the velocity distributions of all hadrons and their 
resonances in the box (or cell), it is more appropriate to confirm 
this isotropy for very few main hadron species, such as nucleons and 
pions, and then examine the isotropy of the integral distribution, 
such as pressure.

Since the number of particles is not a conserved quantity in strong
interactions, the equilibrium in the system cannot be reached until
abundances of all hadron species become saturated. Also, inelastic
collisions affect the thermal distributions of particles. Therefore, 
criteria (i),(ii) should be completed by the requirement of time 
independence of (iii) energy distributions of particles (thermal 
equilibrium) and (iv) their yields (chemical equilibrium). Elastic 
scattering of particles drives the system to {\it thermal\/} 
equilibrium, while inelastic reactions push the system towards the 
stage of {\it chemical\/} equilibrium.

\begin{figure}[htp]
\centerline{\epsfysize=17cm \epsfbox{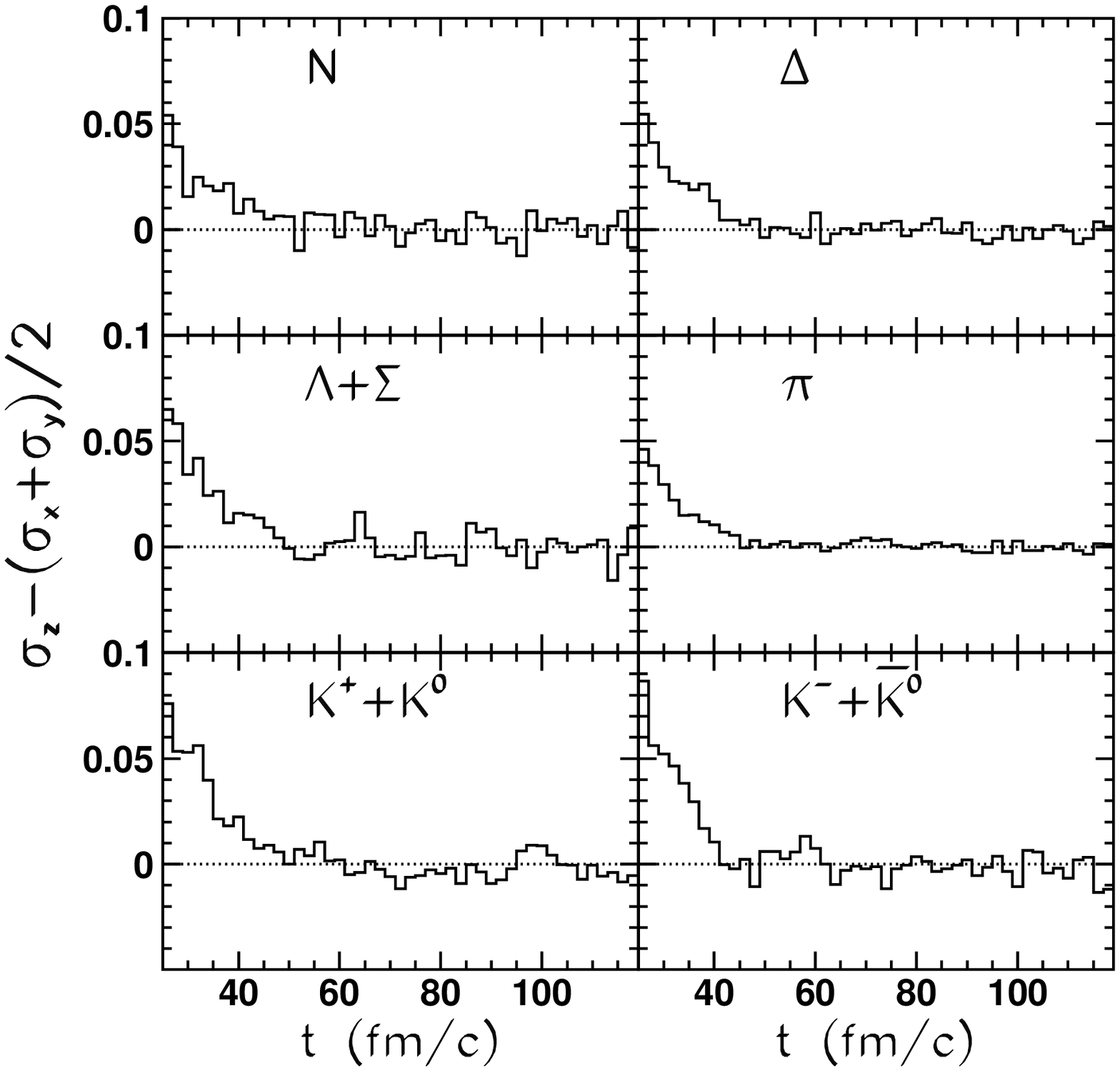}}
\caption{
Anisotropy function 
$f_a^{(2)} = \sigma_z^{(2)} - (\sigma_x^{(2)} + \sigma_y^{(2)}) / 2$
of the velocity distributions of hadron species in the box
with volume $V = 125$ fm$^3$ as a function of time.
Conditions in the box correspond to those of the central 
$V = 125$ fm$^3$ cell in central Pb+Pb
collision at 160{\it A} GeV at $t = 10$ fm/$c$, i.e.,
$\varepsilon = 468$ MeV/fm$^3$, $\rho_{\rm B} = 0.0924$ fm$^{-3}$, 
and $\rho_{\rm S} = -0.00987$ fm$^{-3}$. 
}
\label{fig1}
\end{figure}

\begin{figure}[htp]
\centerline{\epsfysize=17cm \epsfbox{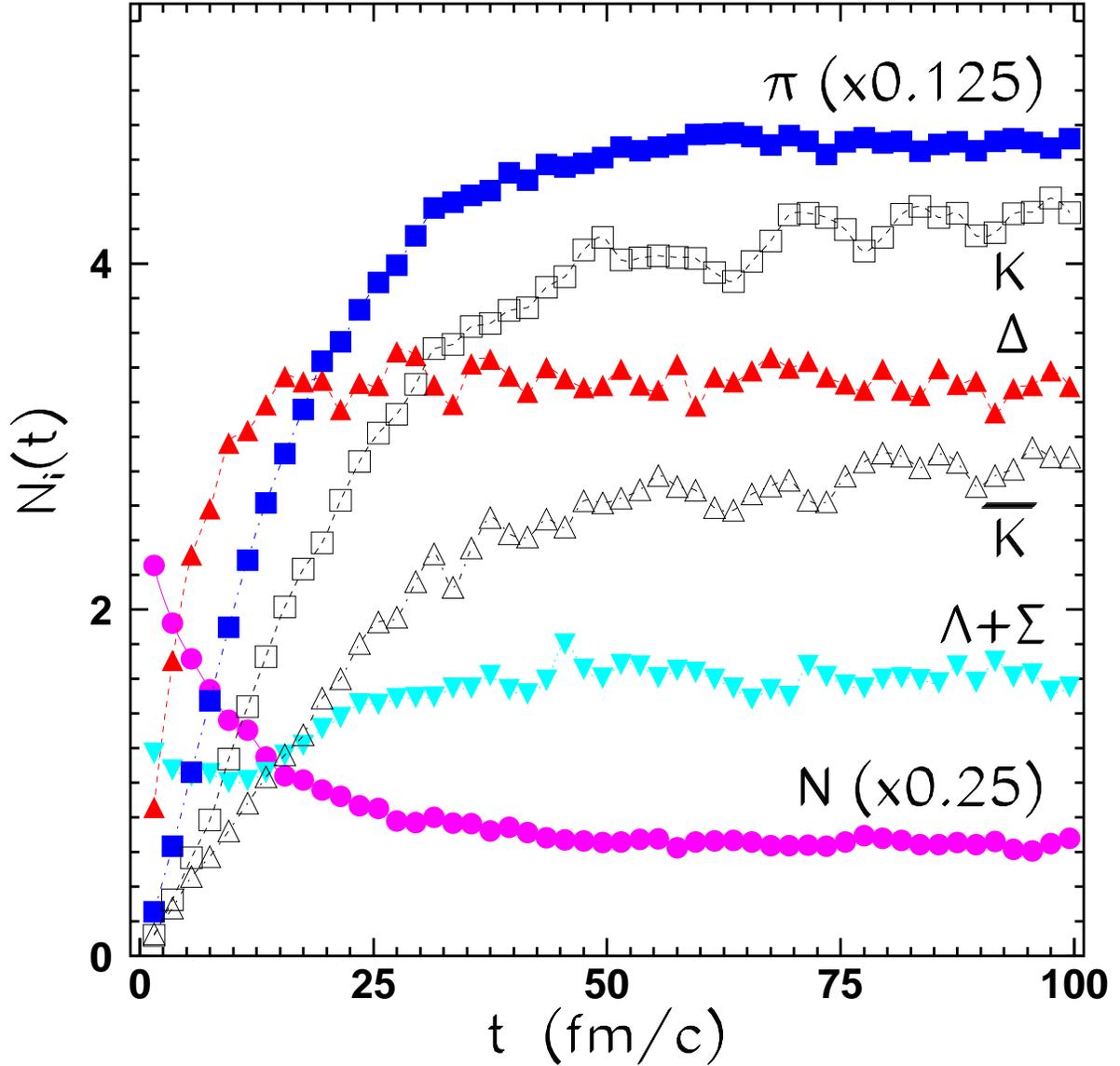}}
\caption{
Time evolution of the yields of $\pi$'s, $K$'s, $\overline K$'s,
$N$'s, $(\Lambda + \Sigma)$'s and $\Delta$'s in the box with
$V = 125$ fm$^3$, $\varepsilon = 468$ MeV/fm$^3$, $\rho_{\rm B} = 
0.0924$ fm$^{-3}$, and $\rho_{\rm S} = -0.00987$ fm$^{-3}$.
}
\label{fig2}
\end{figure}

To examine the course of equilibration in the box the system of
nucleons has been initialized. The values of the energy density, 
baryon density and strangeness density in the box are corresponding 
to those in the central cell with volume $V = 125$ fm$^3$ in Pb+Pb 
central collision at SPS ($E_{\rm lab} = 160${\it A} GeV) energy at 
time $t = 10$ fm/$c$ after the beginning of the collision, namely,
$\varepsilon = 468$ Gev/fm$^3$, $\rho_{\rm B} = 0.09$ fm$^{-3}$,
and $\rho_{\rm S} = -0.01$ fm$^{-3}$. At that time the hadronic 
matter in the cell is already in kinetic equilibrium 
\cite{lep1,lv99jpg}. As seen in Fig.~\ref{fig1} the second moments 
\beq
\ds
\sigma_{x,y,z}^{(2)} = \langle v_{x,y,z}^2 \rangle  - 
\langle v_{x,y,z} \rangle ^2 
\label{eq2a}
\eeq
of the longitudinal and transverse 
velocity distributions of six main hadron species become isotropic 
after 50 fm/$c$. However, some of the hadron abundances are not frozen 
yet. Figure~\ref{fig2} depicts the
evolution of the yields of $\pi$'s, $K$'s, $\overline K$'s, 
$\Delta$'s, $(\Lambda + \Sigma)$'s, and $N$'s with time $t$. The
saturation of the kaon yields is reached at about 70$-$90 fm/$c$ 
after the beginning of the process, while the yields of other hadrons 
are saturated at 50$-$60 fm/$c$.
These saturation times agree with the results of Ref. \cite{Brat00},
where the delay in saturation of the abundances of strange particles
has been also observed.

\begin{figure}[htp]
\centerline{\epsfysize=17cm \epsfbox{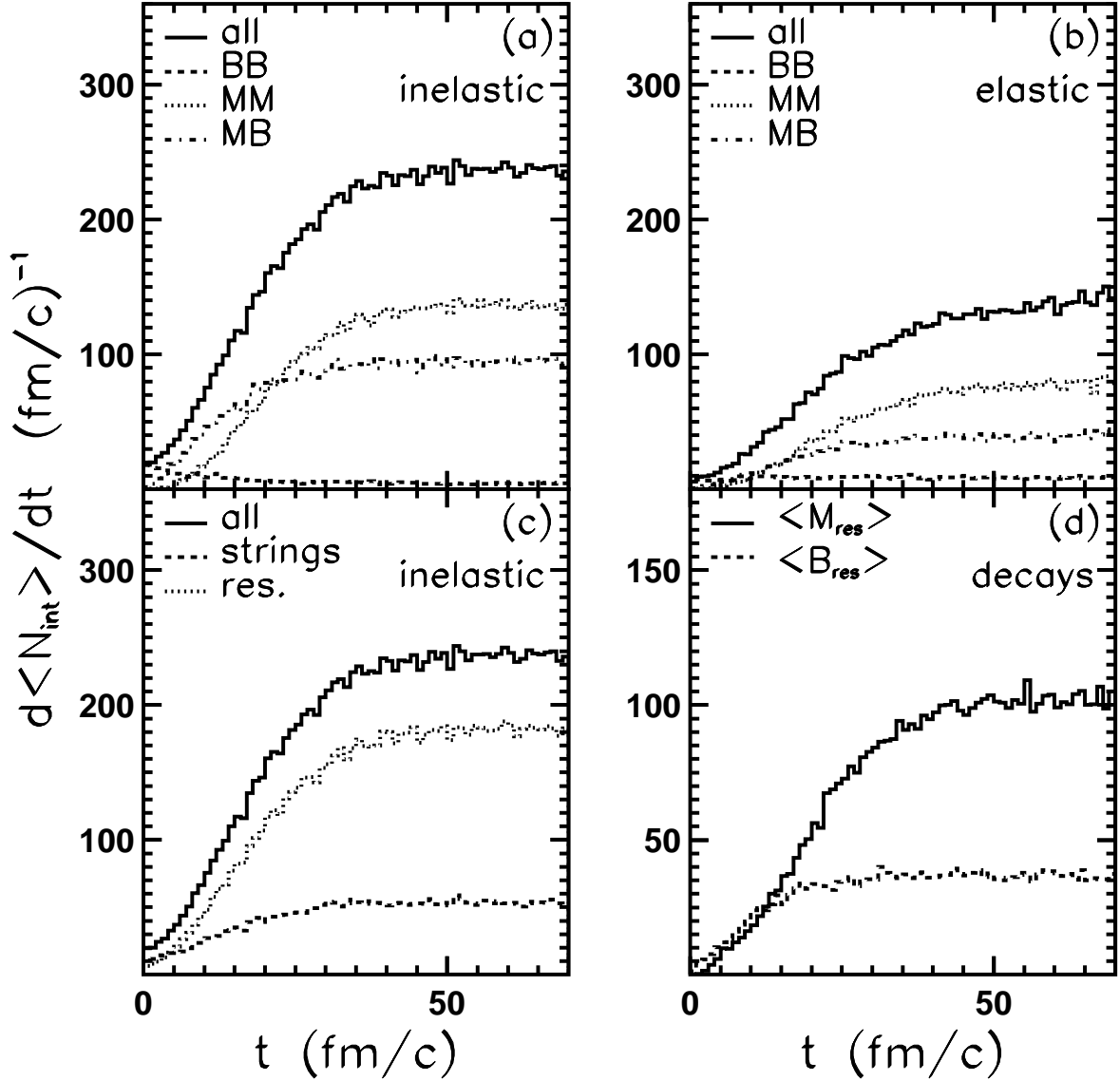}}
\caption{
Time evolution of the number of inelastic (a),(c) and elastic (b) 
collisions, and decays of resonances (d) in the box  with
$V = 125$ fm$^3$, $\varepsilon = 468$ MeV/fm$^3$, $\rho_{\rm B} = 
0.0924$ fm$^{-3}$, and $\rho_{\rm S} = -0.00987$ fm$^{-3}$. 
}
\label{fig3}
\end{figure}

The stationary stage, attained by the hadron matter in the box,
is also characterised by a saturation of the number of inelastic and
elastic collisions of hadrons and decays of resonances per unit of
time. Figure \ref{fig3} shows that about 45\% of all reactions in
the system are inelastic collisions. Another 27\% of interactions
are elastic, and the fraction of decays of resonances is about 28\%.
Compared with the number of resonance decays it means, particularly, 
that the rate of collisions is high and, therefore, more than 30\% of
all resonances suffer elastic or inelastic
collisions with other hadrons within their lifetimes. According
to Fig.~\ref{fig3} both elastic and inelastic reactions are 
controlled by meson-baryon (MB) and meson-meson (MM) interactions
with small admixture of baryon-baryon (BB) collisions. 
Among the inelastic collisions the processes going via the formation
of resonances (78\% of the inelastic cross section) dominate over the
reactions with the string formation (22\% of $\sigma_{\rm inel}$), as 
seen in Fig.~\ref{fig3}(c). 

At the initial nonequilibrium stage the collisions of hadrons are 
very energetic. Then, due to multiparticle processes, the kinetic 
energy of primordial nucleons is converted into the mass of newly 
produced particles, mainly mesons. As a result, the average energies 
of MM, MB and BB interactions, both elastic and inelastic, rapidly 
drop. The mean energy, $\langle \sqrt{s} \rangle$, of elastic and 
inelastic hadron collisions in the box at $t \geq 100$ fm/$c$ is
listed in Table~\ref{tab1}. The value of $\langle \sqrt{s} \rangle$
for elastic and inelastic collisions varies from 0.8 GeV for MM
interactions up to 2.7 GeV for heavy BB systems. At these energies 
the contribution of elastic processes to the total cross section of 
a given reaction is indeed about 40\% smaller than the inelastic 
cross section \cite{pdg96}.

\begin{figure}[htp]
\centerline{\epsfysize=17cm \epsfbox{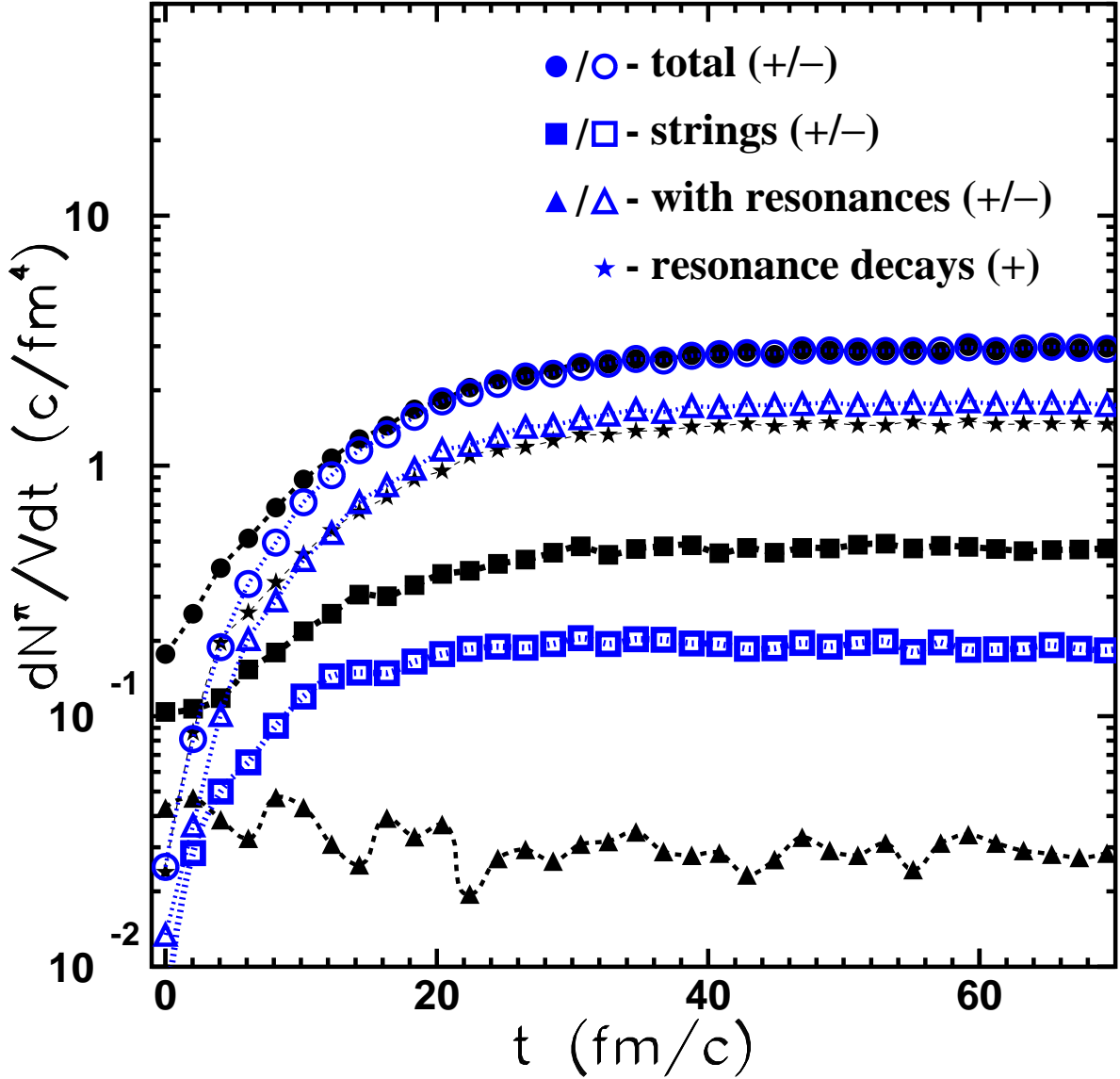}}
\caption{
Production rates for pions in the box  with
$V = 125$ fm$^3$, $\varepsilon = 468$ MeV/fm$^3$, $\rho_{\rm B} = 
0.0924$ fm$^{-3}$, and $\rho_{\rm S} = -0.00987$ fm$^{-3}$. 
Pions are produced (full 
symbols) and absorbed (open symbols) in various inelastic processes, 
including break-up and formation of strings (boxes), and reactions
with resonance excitations (triangles). Pions coming directly from 
the decays of resonances are shown by stars. 
}
\label{fig4}
\end{figure}

Figure~\ref{fig4} clarifies the kinetics of the equilibration of the
pion yields, whose number can only be changed in inelastic reactions. 
The excess of pions produced in the fragmentation of strings is 
compensated by the absorption of pions by resonance matter, e.g., in 
a two-step process such as $\pi \pi \rightarrow \rho$ and 
$\rho \rho \rightarrow \pi \pi$. As mentioned before, because of the 
large particle density and large interaction cross section the 
resonances can interact immediately after their production without 
decay but, probably, with the formation of new strings. Thus the 
symmetry between the gain and loss terms for pions (as well as for 
other hadrons) is restored. Note, that although the chemical 
composition of the hadrons is frozen, the inelastic reactions in the 
system are not completely ceased.

\begin{figure}[htp] 
\centerline{\epsfysize=17cm \epsfbox{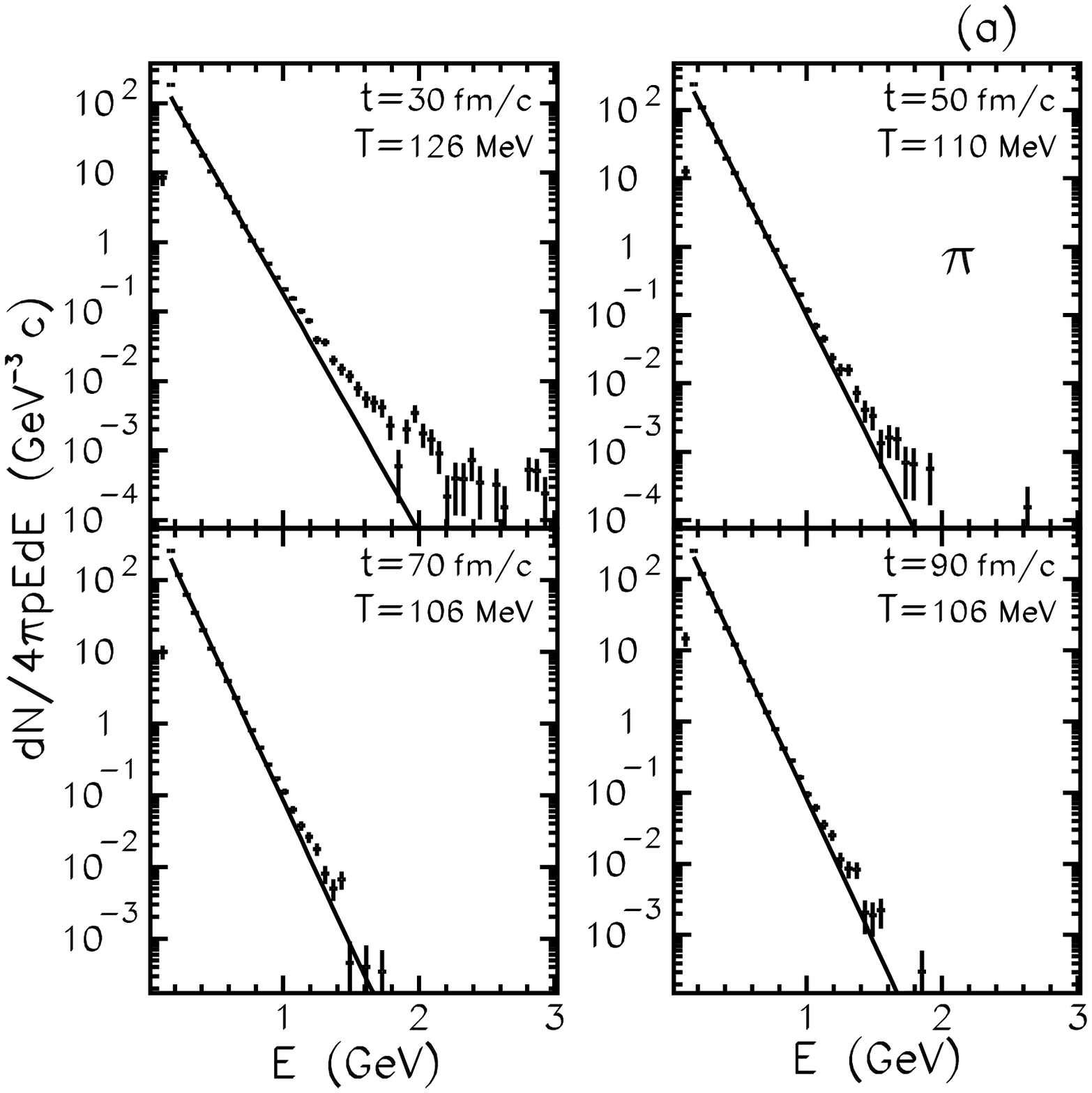}}
\caption{
(a) Time evolution of the energy spectra of pions in the box  with
$V = 125$ fm$^3$, $\varepsilon = 468$ MeV/fm$^3$, $\rho_{\rm B} = 
0.0924$ fm$^{-3}$, and $\rho_{\rm S} = -0.00987$ fm$^{-3}$.
Solid lines are the results of Boltzmann fit to the distributions.
The value of the inverse slope parameter, $T$, is listed in each
panel of the figure.
(b) The same as (a) but for nucleon energy spectra in the box.
}
\centerline{\epsfysize=17cm \epsfbox{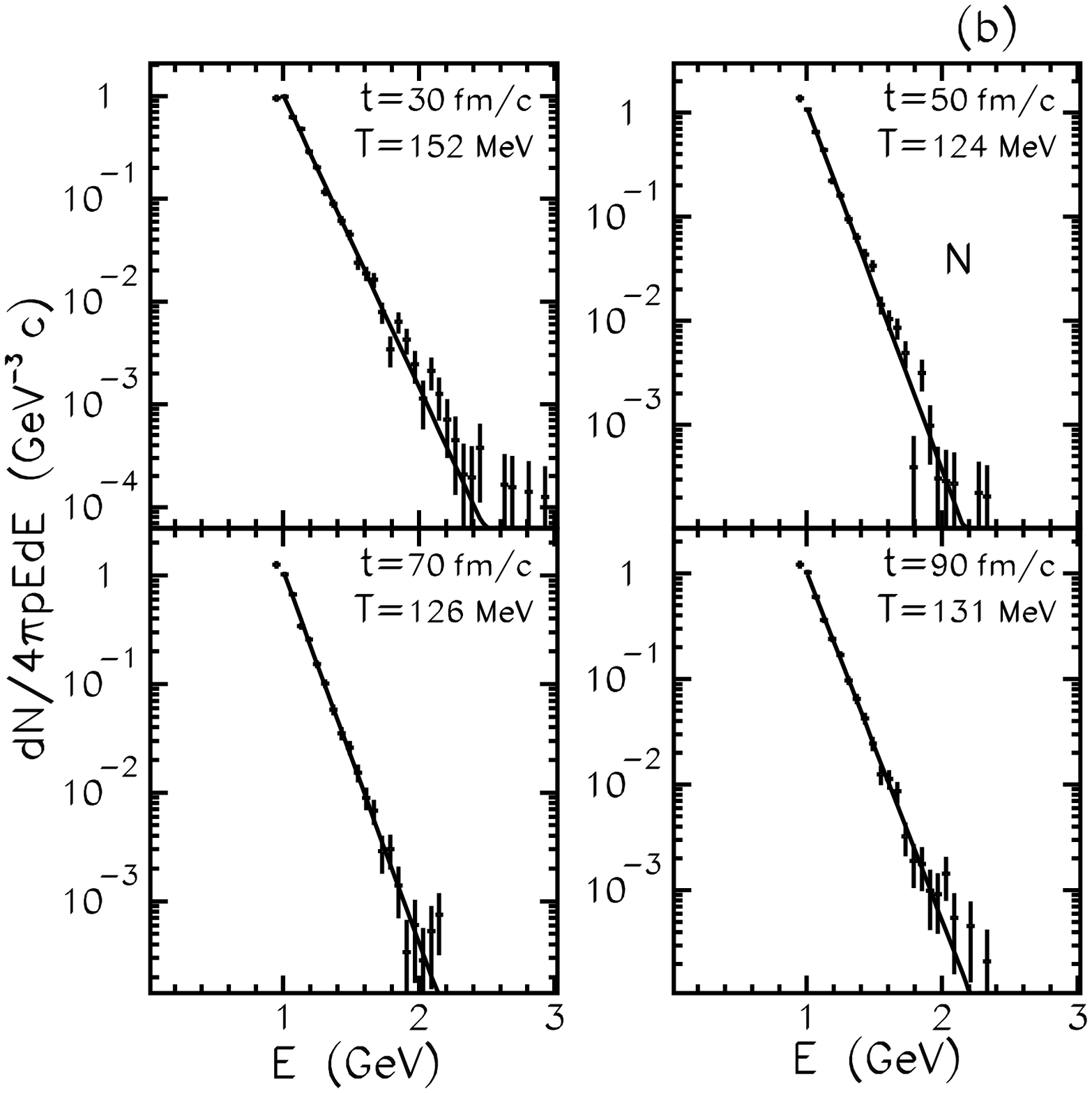}}
\label{fig5}
\end{figure}

To verify that not only chemical but also thermal equilibrium is
attained the energy spectra of hadrons in the box have been studied.
If the system is in thermal equilibrium these spectra should be
nicely fitted to Gibbs distribution, $\exp{ (-E_i/T + \mu_i/T)}$.
Pion energy spectra, $d N / 4 \pi p E d E$, are shown at different 
times in Fig.~\ref{fig5}(a), and nucleon energy spectra are 
represented in Fig.~\ref{fig5}(b). Both spectra are remarkably close 
to the equilibrium fit. Nucleon energy spectra are time independent 
already after $t = 50$ fm/$c$ with the ``apparent" temperature 
$T_{N} \cong 127$ MeV, and the pion energy spectra become
unchangeable at $t \approx 70$ fm/$c$, with the ``apparent" 
temperature $T_{\pi} \cong 106$ MeV. The difference in $T_{\pi}$ and 
$T_N$ looks not very significant, 
$(T_N - T_{\pi}) / T_N \approx 0.17 $. The hadronic matter in the box 
can be considered as a stationary system almost in equilibrium, since 
the conditions (i)$-$(iv) are satisfied.
 
\section{Comparison between cell and box results}
\label{sec3}

\subsection{ Energy spectra and yields of hadrons}
\label{sub3a}

To compare the hadron yields and energy spectra in the central cell 
of heavy-ion collisions at energies 10.7 and 160{\it A} 
GeV with the hadronic spectra in equilibrium, the values of 
$\varepsilon$, $\rho_{\rm B}$, and $\rho_{\rm S}$ have been extracted 
from the cell conditions at each time step. In Ref. \cite{lep1} these
values are substituted into the system of nonlinear equations of
the statistical model to obtain temperature $T$, baryon chemical
potential $\mu_{\rm B}$, and strangeness chemical potential
$\mu_{\rm B}$. The procedure enables to calculate particle 
multiplicities and energy spectra in the SM. 

\begin{figure}[htp] 
\centerline{\epsfysize=15cm \epsfbox{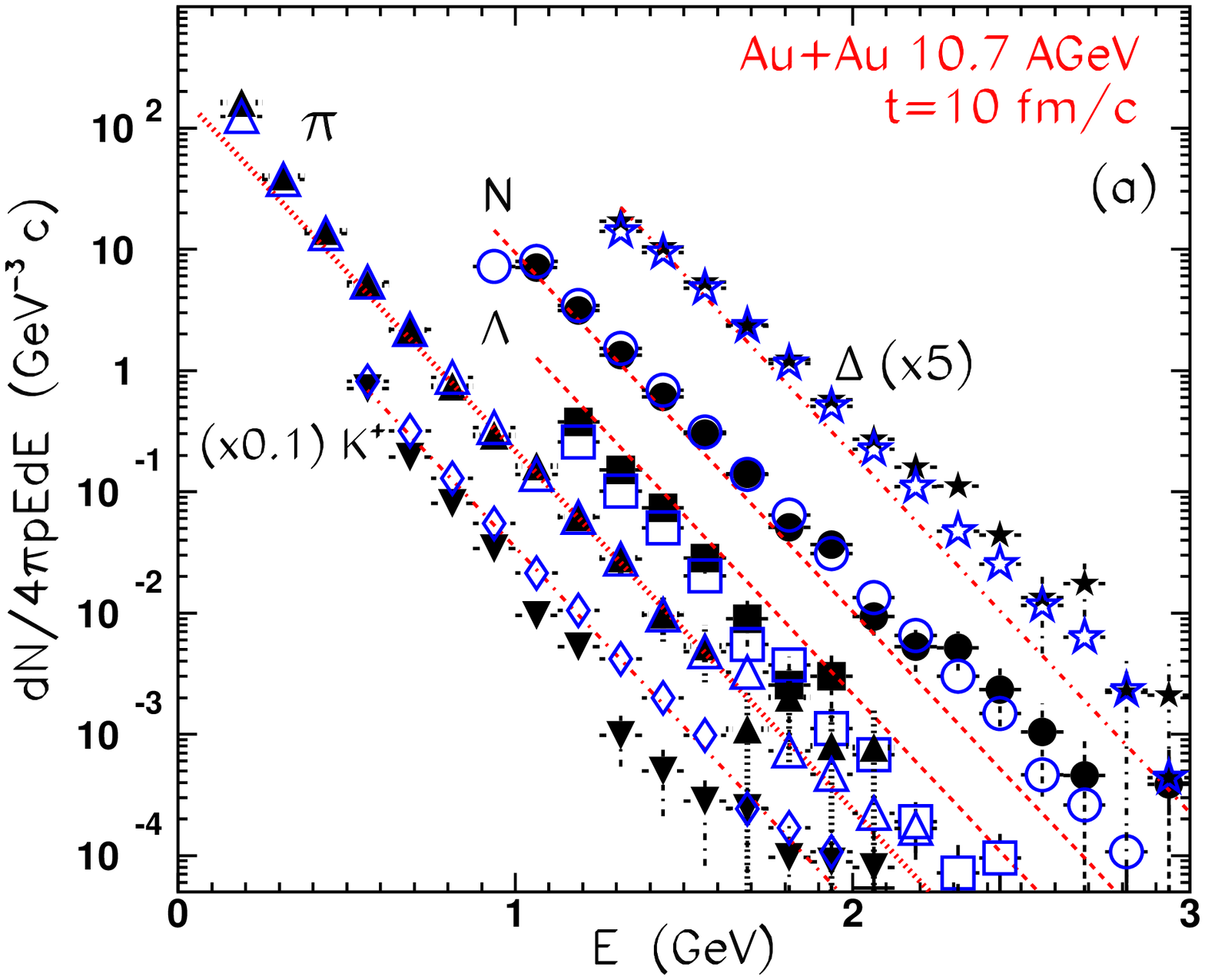}}
\caption{
(a) Energy spectra of $N$ (circles), $\Lambda$ (boxes),
$\pi$ (triangles up), $K^+$ (triangles down and diamonds), 
and $\Delta$ (stars)
in the central 125 fm$^3$ cell of Au+Au collisions at 10.7{\it A} GeV 
at $t$=10~fm/$c$.
Open symbols indicate UrQMD box calculations.
Lines are the results of Boltzmann fit to the distributions
with the parameters $T$=147 MeV, $\mu_B$=510 MeV, $\mu_S$=129 MeV
obtained in the ideal hadron gas model.
(b) The same as (a) but for Pb+Pb collisions at 160{\it A} GeV.
In addition, spectra of $K^-$'s (asterisks and open crosses) are 
plotted. Parameters of the Boltzmann fit are $T$=161 MeV,
$\mu_B$=197 MeV, $\mu_S$=36.8 MeV.
}
\centerline{\epsfysize=15cm \epsfbox{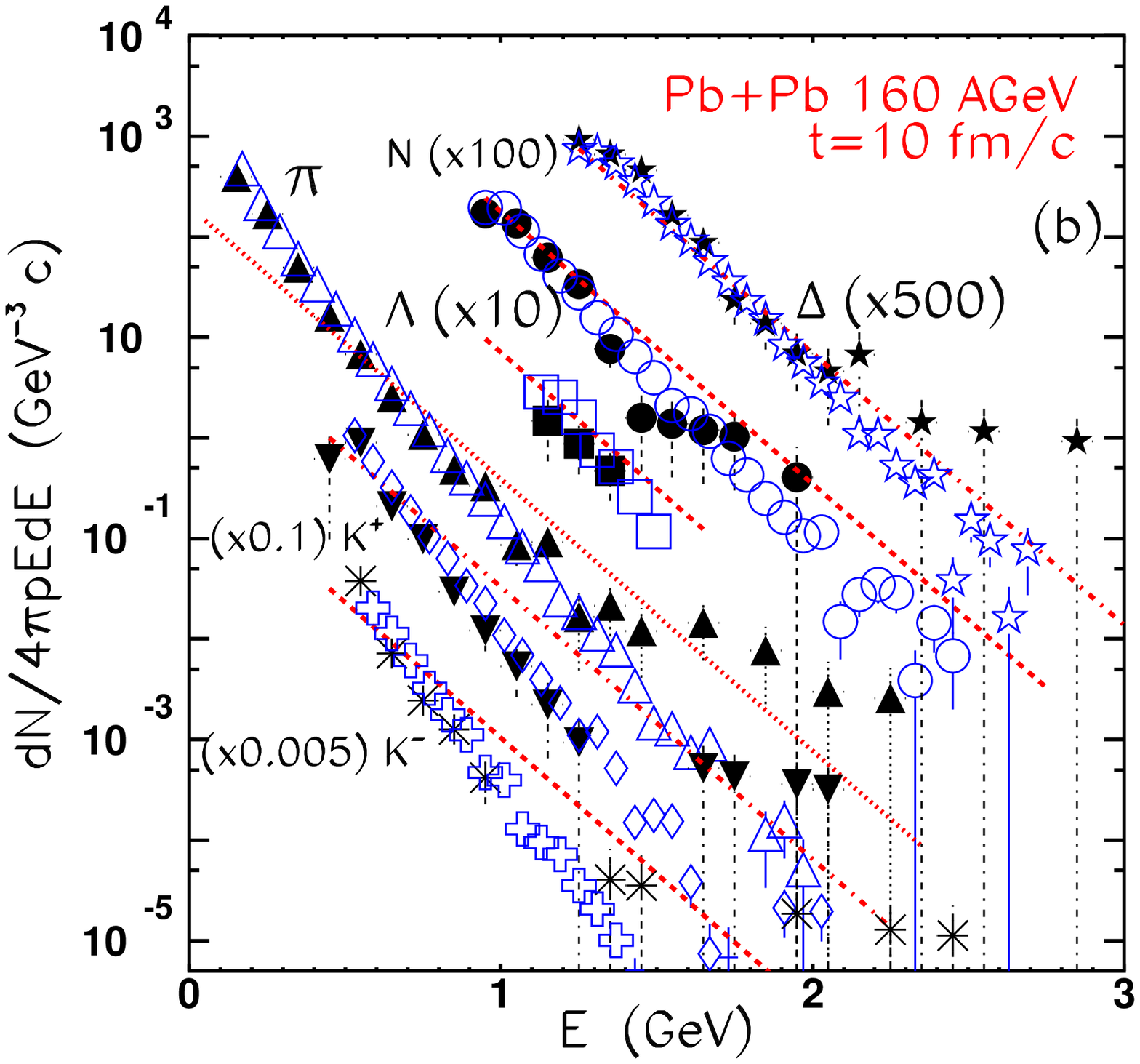}}
\label{fig6}
\end{figure}

In the present analysis the extracted parameters $\varepsilon, 
\rho_{\rm B}, \rho_{\rm S}$ are used to initialize the system of
hadrons in the box. The hadrons start to interact, elastically and
inelastically, and the process of relaxation to the equilibrium
begins. When the fluctuations of hadron yields and their energy
spectra from some average values become small, and the average
values themselves become time independent, the simulations are
stopped. Then, the spectra of hadrons in the cell are compared with 
those of the infinite hadron matter. Energy spectra of hadron species
in the cell at $t = 10$ fm/$c$ are shown in Figs.~\ref{fig6}(a) and 
\ref{fig6}(b) together with the box simulations for both energies.
Predictions of the statistical model of an ideal hadron gas
\cite{Belk98} are plotted as well. We see that 
the agreement between the cell and the box results is very good.
It is insensitive to the rise of bombarding energy from AGS to SPS,
in contrast to noticeable deviations from the predictions of SM.
Values of the inverse slope parameter, extracted from the Boltzmann
fit to particle spectra in the cell and in the box, and temperatures,
given by the SM calculations, are listed in Table~\ref{tab2} for
the central cell in Pb+Pb collisions at 160{\it A} GeV. The thermal
characteristics of hadron species in the cell at $t \geq 10$ fm/$c$
are close to those of the infinite nuclear matter, and rather far
from the ideal equilibrium characteristics (see also Ref. 
\cite{lep1}).

\begin{figure}[htp]  
\centerline{\epsfysize=17cm \epsfbox{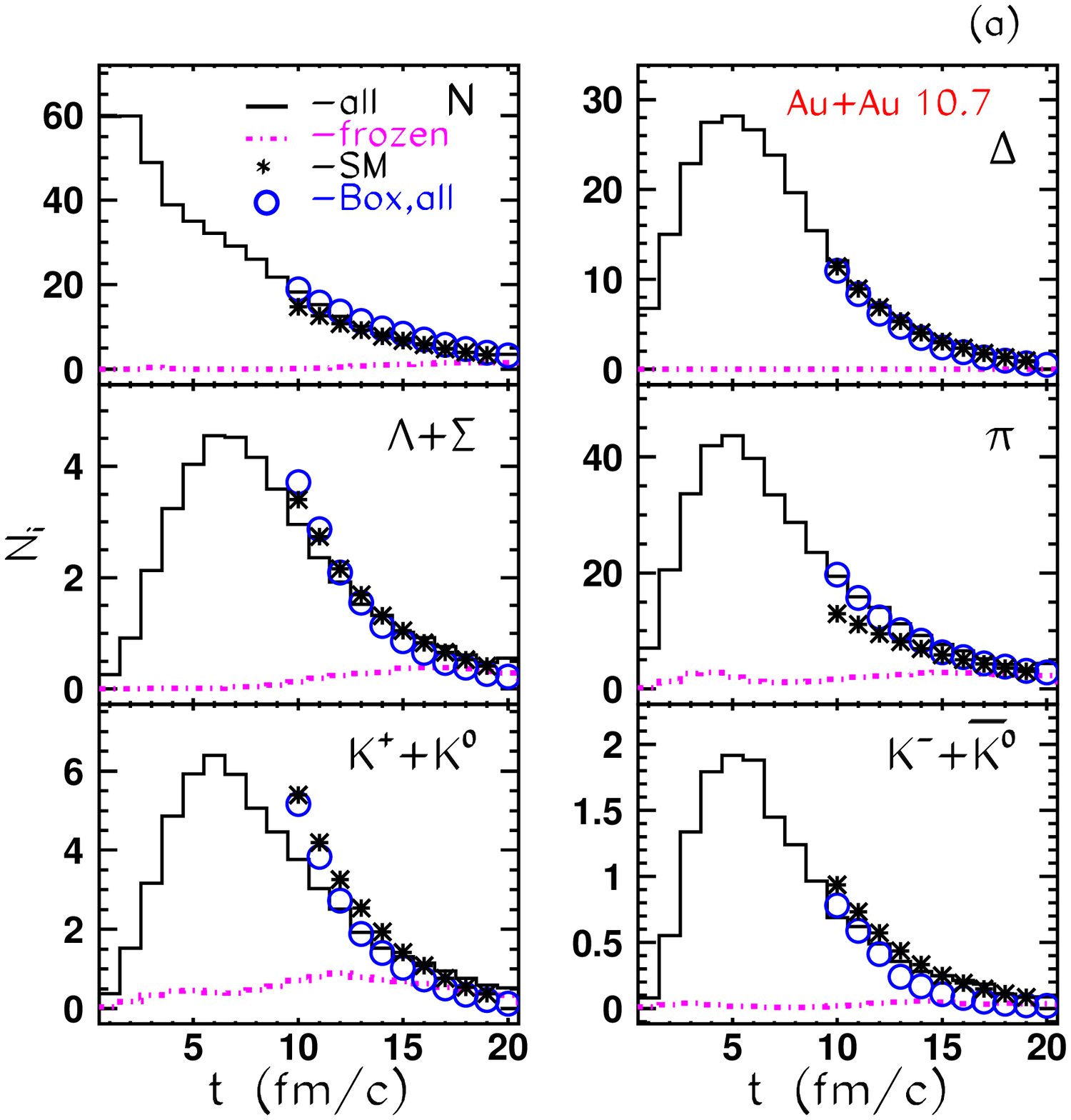}}
\caption{
(a) The number of particles in the central cell of heavy-ion 
collisions at 10.7{\it A} GeV as a function of time as obtained in 
the UrQMD model (histograms) together with the predictions of the 
SM (asterisks) and with the box calculations (open circles). 
The fractions of frozen particles in the cell are shown by 
dot-dashed lines.
(b) The same as (a) but for Pb+Pb collisions at 160{\it A} GeV.
}
\centerline{\epsfysize=17cm \epsfbox{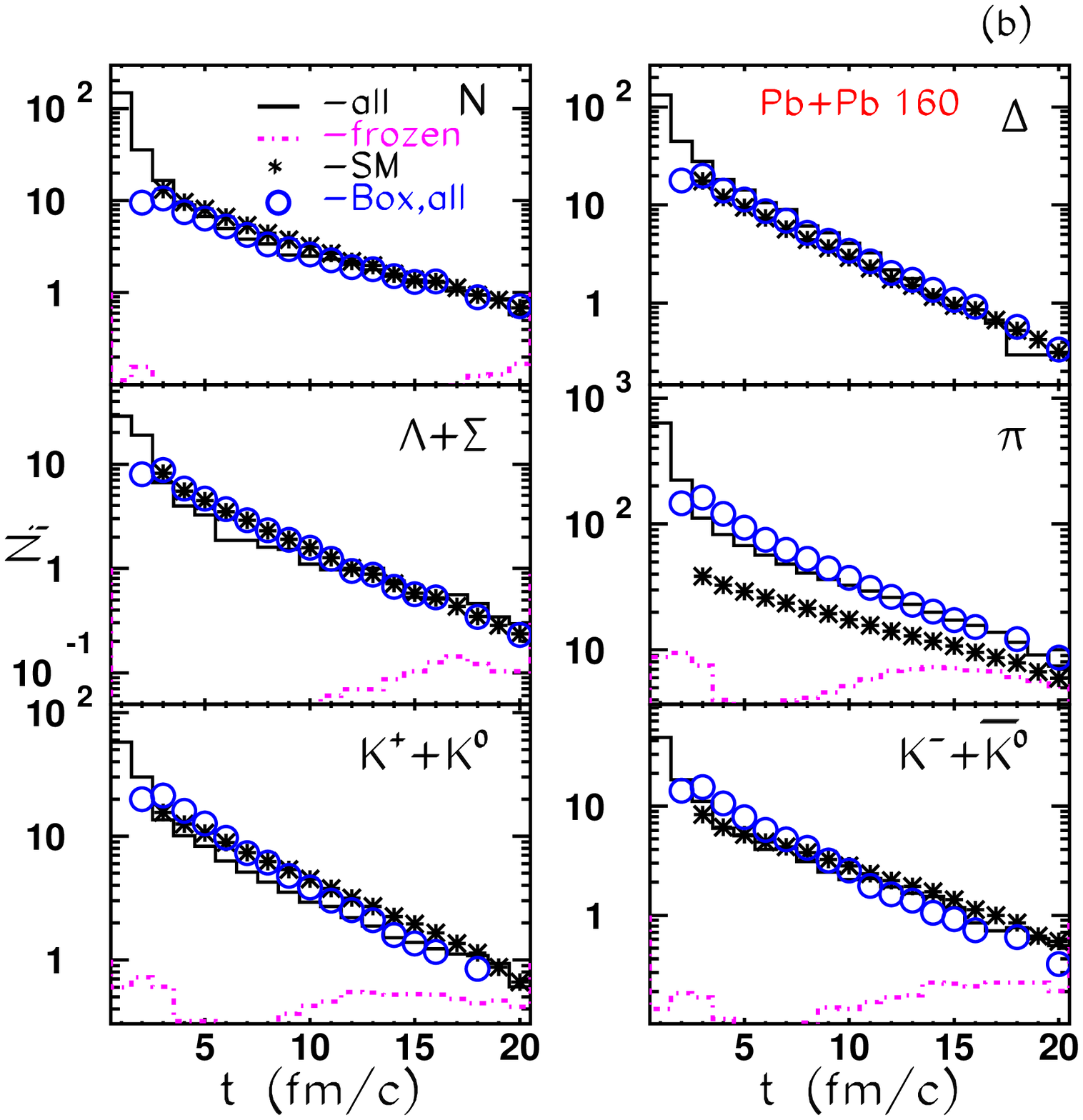}}
\label{fig7}
\end{figure}

The abundances of hadrons in the cell for both reactions are 
presented in Figs.~\ref{fig7}(a) and \ref{fig7}(b). The fractions
of particles belonging to the final state of the heavy-ion collisions 
(``frozen" particles) are plotted onto the results also. Again, the 
yields of hadrons in the cell at $t \geq 10$ fm/$c$ are close to the 
corresponding yields in the box for all species. Statistical model 
underestimates the number of pions, especially at high-energy 
densities, but provides a better agreement for the baryon sector. 
One can conclude that the hadron matter in the cell reaches at the
late stage a quasiequilibrium state, similar to the state of infinite 
hadron matter simulated within the same microscopic model. This state 
is dubbed steady state \cite{KoPr98,irr99}. On the 
other hand, the steady state does not match to the ideal thermal and 
chemical equilibrium, i.e., the state of maximum entropy. It means, 
particularly, that the conditions (i)$-$(iv) formulated
in Sec.~\ref{sec2} are the necessary and sufficient criteria of the
stationary (or steady) state only. Fulfilment of these conditions 
cannot guarantee thermal and chemical equilibration in the system of 
interacting hadrons from the point of view of equilibrium statistical
mechanics. 

The controversial point of the discussion is that the
entropy of a statistical system should have a pronounced peak
corresponding to the equilibrium configuration \cite{LaLi80}.
The experimentally measured mean multiplicities of secondaries in 
hadronic and nuclear collisions seem to agree roughly with the
estimations of the SM, which links the multiplicity of the produced
particles to the entropy of the thermalized system. Thus, the 
entropy of the steady state cannot be much smaller than the entropy
given by the statistical model. To resolve the ambiguity it is 
important to study how much the entropy density in the cell 
deviates from the maximum entropy density allowed to the system.

\subsection{Entropy analysis}
\label{sub3b}

From here we will use $s$ to denote the entropy density. For the 
statistical system of volume $V$ in equilibrium the entropy density 
$s = S/V$ can be determined from the Gibbs thermodynamical identity
\beq
\ds
T s =  \varepsilon - \mu_{\rm B} \rho_{\rm B} -
\mu_{\rm S} \rho_{\rm S} + P \quad,
\label{eq3}
\eeq
where $P$ denotes the pressure in the system. The alternative way
is to define the entropy density via the particle distribution
function $f(p,m_i)$:
\beq
\ds
s = -\sum_i \frac{g_i}{\left( 2\pi\hbar \right)^3} \int_0^{\infty}
   f_i(p,m_i)\, \left[ \ln{f_i(p,m_i)}-1 \right] \, d^3 p \quad,
\label{eq4}
\eeq
with $g_i$ and $m_i$ being the degeneracy factor and mass of the
hadron species $i$, respectively. The last expression permits one to 
calculate the entropy density by means of the microscopic 
distribution function
\beq
\ds
f_i^{\rm mic}(p) = \frac{(2\pi \hbar)^3}{V g_i} \frac{dN_i}{d^3 p} 
\quad.
\label{eq5}
\eeq

\begin{figure}[htp]  
\centerline{\epsfysize=17cm \epsfbox{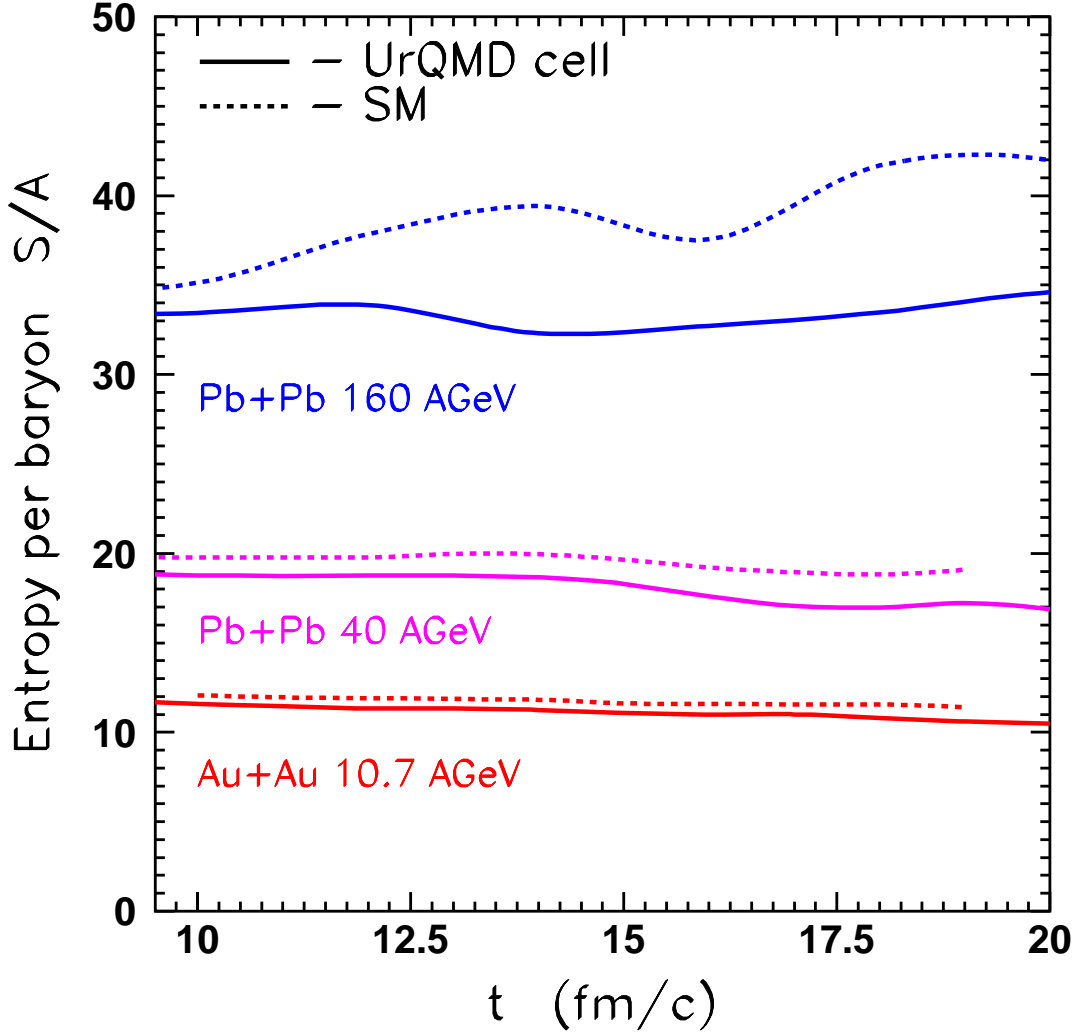}}
\caption{
Time evolution of the entropy per baryon ratio, $S/A$, in the central
cell with $V = 125$ fm$^3$ in heavy-ion collisions at 10.7, 
40, and 160{\it A} GeV. Solid lines indicate microscopic 
calculations with the UrQMD model. Dashed lines show the predictions 
of the SM of an ideal hadron gas, obtained with the same values of 
baryon density, energy density and strange density, as extracted from 
the analysis of the cell conditions in the UrQMD simulations.
}
\label{fig8}
\end{figure}

If the hadron cocktail reaches thermal and chemical equilibrium, the
distribution functions of hadrons are given simply by 
Bose-Einstein or Fermi-Dirac distributions. In this case both 
Eqs.~(\ref{eq3}) and (\ref{eq4}) should provide the same results 
for $s$. But: the hadron matter in the central cell (and in the
box) is neither in chemical nor in thermal equilibrium at 
high energy densities. Therefore, the entropy densities calculated by 
Eq.~(\ref{eq4}) should be smaller than those predicted by the 
statistical model Eq.~(\ref{eq3}). This difference is clearly seen
in Fig.~\ref{fig8}, where the entropy densities per baryon, $S/A$,
in the central cell at the quasiequilibrium stage are compared 
with the SM calculations. The $S/A$ ratio in the cell is nearly
constant during the late evolution of the system. Its deviation from 
the equilibrium value is about 4\% at AGS and about 6$-$10\% at
SPS energy.

\begin{figure}[htp]
\centerline{\epsfysize=17cm \epsfbox{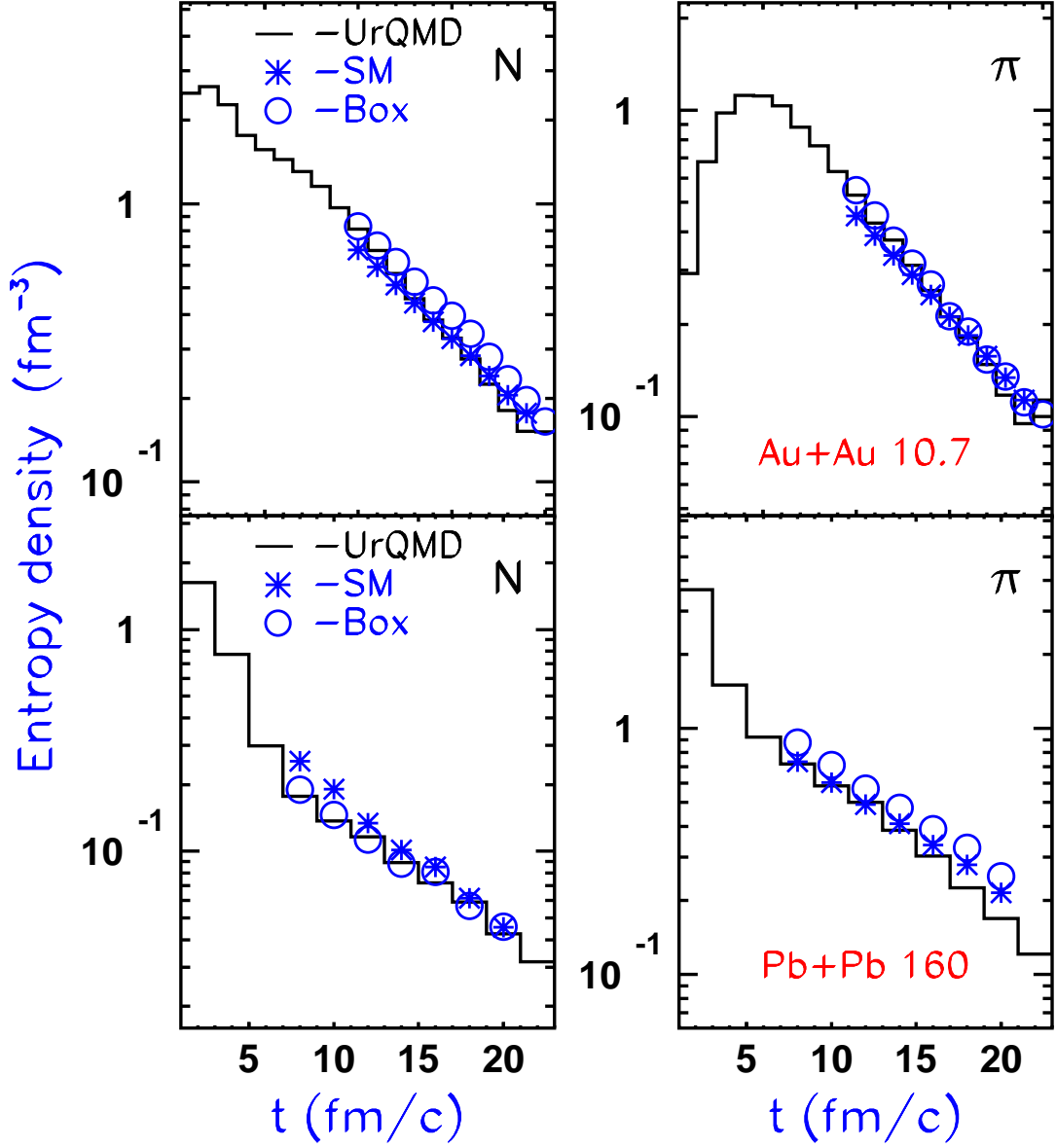}}
\caption{
Partial entropy densities of $N$'s and $\pi$'s in the central cell of 
heavy-ion collisions at 10.7{\it A} GeV (upper panels) and 160{\it A} 
GeV (lower panels). Histograms denote the UrQMD simulations. 
Predictions of the SM and box calculations are shown by asterisks
and by open circles, respectively.
}
\label{fig9}
\end{figure}

Partial entropy densities carried by nucleons and by pions in the 
central cell are shown in Fig.~\ref{fig9} together with the box
and SM calculations. The entropies of hadron species in the 
cell appear to be very close to the equilibrium values, predicted 
by the SM, despite the difference in hadron abundances. For
instance, although the number of pions in Pb+Pb collisions at SPS
energy is twice as large in the cell at $t=10$ fm/$c$ as in the
ideal equilibrium [see Fig.~\ref{fig7}(b)], the partial entropy
densities are nearly the same, $s_\pi^{\rm mic} \cong s_\pi^{\rm SM}$. 
This is due to the fact that the apparent temperature of pions in the 
cell is much lower than the SM temperature listed in Table~\ref{tab2}.
Therefore, the total entropy of the steady state, attained by the
hadron matter in the central cell or by the infinite hadron matter, 
is close to the maximum entropy, assigned to the system in thermal
and chemical equilibrium. 

\section{Which equilibrium is true?}
\label{sec4}

In Ref. \cite{lep1} it was shown that the hadronic yields and energy 
spectra in the central cell of relativistic heavy-ion collisions do
not coincide with the corresponding spectra, given by the SM, not even
at the stage of kinetic equilibrium. Formally this means that thermal 
and chemical equilibrium in the cell is not reached yet. For instance,
times $t \leq 20$ fm/$c$ may be too short compared to the typical
equilibration times. 

\begin{figure}
\centerline{\epsfysize=16cm \epsfbox{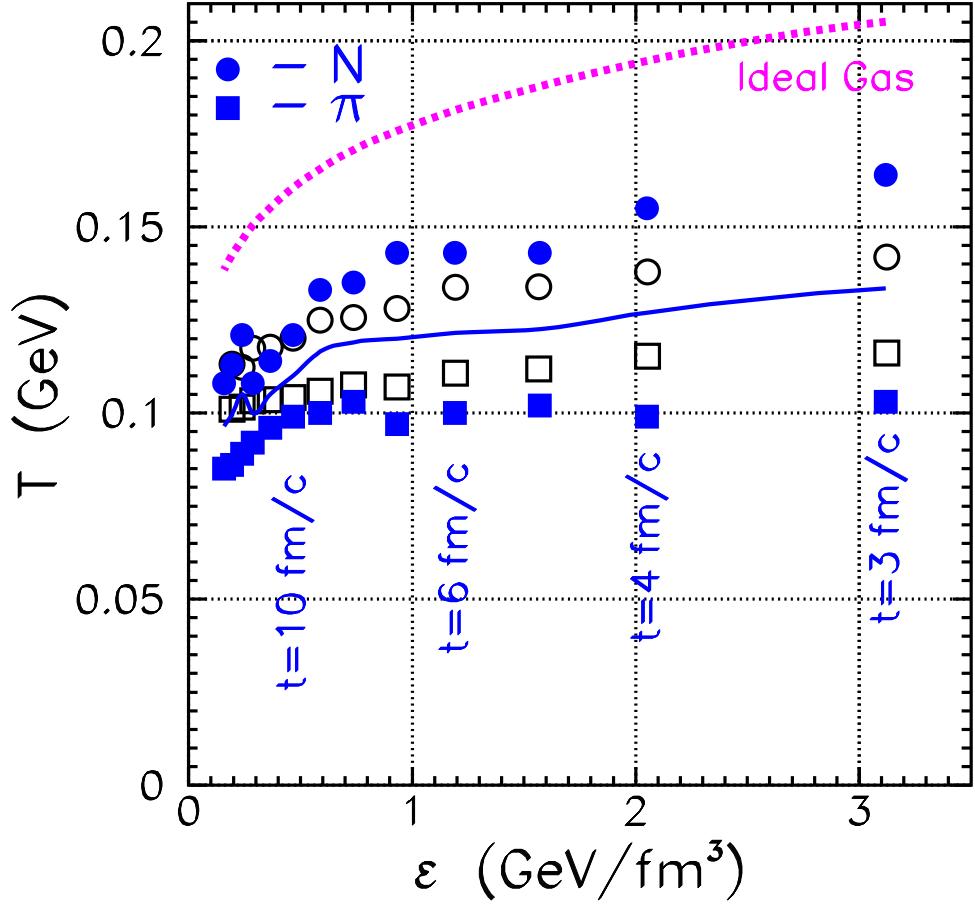}}
\caption{
Time evolution of energy density and temperature in the central cell
of central Pb+Pb collisions at 160{\it A} GeV (full symbols) compared
to the equilibrated infinite hadron matter (open symbols) and to the
predictions of the ideal gas SM (dashed curve) calculated with the
same $\varepsilon$, $\rho_{\rm B}$, and $\rho_{\rm S}$. Temperatures
are extracted as the inverse slopes of Boltzmann fits to the 
energy spectra of nucleons (circles) and pions (squares).
Solid curve indicates the average temperature of hadron mixture in
the cell. 
}
\label{fig10}
\end{figure}

The present study reveals, however, that the hadron distributions in
the cell after $t \cong 10$ fm/$c$ agree well with the hadron spectra 
of infinite hadron matter, simulated within the same microscopic
model. At relatively low energy densities the results of the box 
simulations are close to the predictions of the statistical model.
The accord is broken at energy densities higher than 0.5 GeV/fm$^3$.
Which equilibrium is true and what is the origin of the disagreement? 
To answer these questions it is necessary to analyse the temperatures
of the hadron species in the box as a function of energy density.
Figure~\ref{fig10} presents the UrQMD simulations in the cell and in 
the box together with the SM estimates for Pb+Pb collisions at 
160{\it A} GeV. At the high density phase of a heavy-ion collision the
cell results exhibit dramatic deviations both from the SM and from
the box calculations. After $t \geq 10$ fm/$c$ (cell time only)
the difference between the box and the cell data is small. 
But both microscopic model simulations diverge clearly from the 
predictions of the statistical model. As was noticed in Ref. 
\cite{Belk98}, the $T(\varepsilon)$ dependence in the UrQMD box is the 
same as in the SM if the multiparticle processes $2 \rightarrow N\ (N 
\geq 3)$ and the many-body decays of resonances have ceased in the 
system. This explains the agreement between the two models at 
low-energy densities. At high energy densities the microscopic model 
predicts the appearance of limiting temperatures similar to that of 
the statistical bootstrap model \cite{SBM} for each of the hadron 
species. These temperatures lie within the range 105 MeV 
$\leq\ T_{\rm lim}\ \leq$ 145 MeV \cite{lep1,Belk98,Bran99}. 
Slightly higher limiting temperature $T_{\rm lim} = 150 \pm 5$ MeV 
has been found in Ref.  \cite{Brat00}. The deviations in model 
calculations can be attributed to a different number of hadron 
species, which is higher in the UrQMD, and to different threshold of 
the excitation of strings \cite{Brat00}. Although the temperatures of 
hadrons are relatively close to each other, the whole system is 
not characterised by a unique temperature. Therefore, the system is
not in the conventional thermal equilibrium.

The chemical composition of the hadron system in the box freezes
after a while (see Sec.~\ref{sec2}). The excess of hadrons, mainly 
pions, produced in inelastic collisions is absorbed by the resonance 
matter. For instance, the reaction $N\pi \rightarrow N\pi\pi$ may be 
compensated by a two-step process $N\pi \rightarrow \Delta$ and 
$\pi \Delta \rightarrow N\pi$ (this is only a simplified example, the 
real number of production and absorption channels is $> 100$). 
Does it mean that the conditions of equilibrium are
fulfilled? The striking answer is {\bf no}, because the 
chemical equilibrium assumes unambiguously that the rates of 
every coupled direct and inverse reaction in equilibrium must be the 
same \cite{LaLi80,Tolm67}. The cyclic process during which the system 
transforms from the quantum state $\cal L$ to quantum state $\cal K$ 
and then comes back to the initial state $\cal L$ via an intermediate 
state $\cal M$ violates the necessary conditions of equilibrium
\cite{Tolm67}. Thus, although the yields of hadrons are time 
independent, the system is not in chemical equilibrium, as defined
by equilibrium statistical mechanics. 

It is important to explain the difference between the evolution
scenarios predicted by the hydrodynamic and microscopic models for
the heavy-ion collisions at ultrarelativistic energies. According
to typical hydrodynamic picture \cite{BGGL92}, the system of hadrons
is produced in chemical and thermal equilibrium shortly after the 
QGP hadronization. Because of the expansion of the system its 
temperature, energy density and particle density drop. The mean
free path for inelastic collisions grows much faster than that for 
the elastic ones. For instance, the mean free path for inelastic
and elastic $\pi\pi$ collisions increases with the power $T^{-9}$
and $T^{-5}$, respectively \cite{GGL90}. Chemical freeze-out takes
place when the mean free path for inelastic collisions becomes 
comparable with the radius of the fireball, i.e., significantly
earlier than the thermal freeze-out. After the (sequential) chemical
freeze-out the system quickly gets out of chemical equilibrium,
heavy resonances decay and chemical potentials start to develop
\cite{BGGL92}. The thermal equilibrium is maintained until the 
thermal freeze-out, when the mean free path of elastic collisions 
rises above the linear dimensions of the system.

The altered scenario is provided by the microscopic model simulations.
Here the system of hadrons in the central cell at $t \geq 10$ fm/$c$
is neither in chemical nor in thermal equilibrium. Despite the fact 
that hadron abundances are frozen, competing inelastic processes
take place still. Also, the temperatures of hadron species are 
different but quite stable, as shown by the simulations of the
infinite nuclear matter. 
 
Although it is still questionable whether or not
the microscopic hadron-string models can adequately describe
the early stage of ultrarelativistic heavy-ion collisions, these
models are really good in description
of the late hadronic phase until the freeze-out of particles.
It is the late stage of heavy-ion collisions that is studied in
the microscopic transport model in the present paper. The rates of
elastic and inelastic processes in the box with the values of
$\varepsilon,\ \rho_{\rm B}$ and $\rho_{\rm S}$ corresponding to
those of the central cell in heavy-ion collisions at AGS and SPS
energies at the stage of kinetic equilibrium are listed in 
Table~\ref{tab3}. The number of processes going via the string 
formation is rather small (less than 10\%) compared to the number 
of elastic reactions and to the number of processes with the 
formation and decay of resonances. It is easy to see that the total 
balance of particles is fulfilled, i.e., 2.47 (34.35) particles per 
fm/$c$ are produced in the decays of resonances and strings, and 
essentially the same 2.47 (34.28) particles per fm/$c$ are absorbed 
by resonance matter in the box with AGS (SPS) cell conditions. The 
mean total energy per hadron, 1.0$\pm$0.1 GeV (see Table~\ref{tab1}), 
is equal to the value 1 GeV of the mean total energy per hadron at
chemical freeze-out, estimated from the statistical model fit to
various experimental data from several GeV up to few hundred GeV
per nucleon (see, e.g., \cite{ClRe99} and references herein).
But the hadron content and the temperatures are far from those given 
by the SM. This difference arise because of the 
lack of inverse reactions, $N \rightarrow 2$, which are not
included in any of the available microscopic models designed for
the description of heavy-ion collisions.

\section{Conclusions}
\label{sec5}

The results may be summarized as follows.
The microscopic transport UrQMD model is employed to study the 
appearance of local equilibrium in the central cell of heavy-ion
collisions at relativistic energies from 10.7 to 160{\it A} 
GeV. The hadron spectra in the cell are compared with the 
corresponding spectra of the infinite nuclear matter, reproduced in 
a box with periodic boundary conditions \cite{Belk98,Bran99} within 
the same microscopic model. A comparison with the predictions of the 
statistical model of an ideal hadron gas has been done in Ref.
\cite{lep1} in detail.

The yields and energy spectra of hadrons in the cell at $t\geq 10$
fm/$c$ are found to be close to those in the box irrespective of 
the c.m. energy of colliding nuclei. The late hadron matter in the 
UrQMD cell, therefore, reaches a quasiequilibrium stage, dubbed steady 
state \cite{KoPr98,irr99}. 
On the other hand, the steady state of hadron matter even at not
very high-energy densities differs from the idealised equilibrium
state assumed by the SM. 

The total entropy density is close to the maximum entropy density 
given by the SM. At SPS energy, for instance, the difference is not 
too large, $(s_{\rm tot}^{\rm SM} - s_{\rm tot}^{\rm mic})/
s_{\rm tot}^{\rm SM} \cong 6$\% only. Moreover, the entropy per 
baryon in the cell remains constant in the time interval 
$10 \leq t \leq 20$ fm/$c$ for energies varying from 10.7 
to 160{\it A} GeV. This explains why the rapidity distribution of 
secondaries in hadronic and nuclear interactions, predicted by 
hydrodynamic models at those energies, fits well to 
the experimental data.
  
However, the hadrons in the UrQMD steady state do not exhibit a 
universal temperature for all species. The yields of mesons, 
especially pions, deviate from the calculations with the SM.
It has been verified previously \cite{lep1} that the assumption of a 
``partial" thermalization of all hadron species except pions cannot 
improve the quality of the fit to the SM. Elimination of pions 
causes the rise of the strangeness chemical potential \cite{lep1}, 
but not the decrease of the temperature of baryons.

The origin of the deviations is traced to different rates of direct 
and inverse reactions for the multiparticle processes and many-body 
decays of resonances, as well as nonzero lifetimes of the 
resonances. 
Absence of the reactions $N \rightarrow 2$ and $N \rightarrow 1\ 
(N \geq 3)$, which is a feature of all microscopic models, leads
to appearance of a cyclic process in the microscopic calculations.
The cyclic process where, e.g., pions are produced in inelastic 
collisions and then are absorbed by resonance matter, violates the
equivalence of rates of every coupled direct and inverse reaction, 
which is the basic principle of the equilibrium statistical
mechanics. 

\section*{Acknowledgments} 
We would like to thank L. Csernai, M. Gyulassy, Amand Faessler, 
and E. Shuryak for the stimulating discussions and
valuable comments.
This work was supported by the Graduiertenkolleg f{\"u}r Theoretische
und Experimentelle Schwerionenphysik, Frankfurt--Giessen, the
Bundesministerium f{\"u}r Bildung und Forschung, the Gesellschaft
f{\"u}r Schwer\-ionen\-for\-schung, Darmstadt, Deutsche
Forschungsgemeinschaft, and the Alexander von Humboldt-Stiftung,
Bonn. S.A.B. acknowledges financial support from the U.S. National
Science Foundation, Grant No. PHY-9605207.

\newpage

\newpage
\mediumtext

\begin{table}
\caption{
The mean total energy, $\sqrt{s}$, of elastic and inelastic collisions
in the box with
$V = 125$ fm$^3$, $\varepsilon = 468$ MeV/fm$^3$, $\rho_{\rm B} = 
0.0924$ fm$^{-3}$, and $\rho_{\rm S} = -0.00987$ fm$^{-3}$ 
at the quasiequilibrium stage. 
Of each pair of
numbers for inelastic collisions the upper one corresponds to the
reaction channel via the string excitations ({\it S\/}), and the lower 
one corresponds to the resonance excitation reactions ({\it R\/}). 
}

\begin{tabular}{clcc}
\multicolumn{2}{c} {\rm Type of reaction} & $N_{\rm react}$ (\%) &
 $ \sqrt{s}$ (GeV) \\
\tableline\tableline
                   & inelas., {\it S}   & 10 & 1.9  \\
   all             & inelas., {\it R}   & 35 & 1.1  \\
 reactions         & elastic, tot.      & 27 & 1.34 \\
  (100\%)          & {\it B} decays     &  8 & 1.36 \\
                   & {\it M} decays     & 20 & 0.81 \\
\tableline
                   & inelas., {\it S}   &  1 & 4.2  \\
baryon-baryon (BB) & inelas., {\it R}   & 32 & 2.7  \\
  (100\%)          & elastic, tot.      & 67 & 2.7  \\
\tableline
                   & inelas., {\it S}   & 10 & 1.6  \\
meson-meson (MM)   & inelas., {\it R}   & 54 & 0.8  \\
   (100\%)         & elastic, tot.      & 36 & 0.91 \\
\tableline
                   & inelas., {\it S}   & 25 & 2.1  \\
meson-baryon (MB)  & inelas., {\it R}   & 46 & 1.4  \\
   (100\%)         & elastic, tot.      & 29 & 2.0  \\
\end{tabular}
\label{tab1}
\end{table}

\begin{table}
\caption{
The temperature $T_{\rm SM}^{\rm all}$ extracted from the SM fit to 
UrQMD data at given $\varepsilon,\ \rho_{\rm B}$, and $\rho_{\rm S}$
(determined from the cell),
together with the temperature of nucleons $T^N_{\rm cell/box}$ and
pions $ T^{\pi}_{\rm cell/box}$ obtained by the Boltzmann fit to 
energy spectra of particles at 160{\it A} GeV within the time interval 
10 fm/$c$ $\leq t \leq 15$ fm/$c$ in the central cell/box.
}

\begin{tabular}{cccccc}
time & $T^{\rm all}_{SM}$ & $ T^N_{\rm cell}$ & $ T^N_{\rm box}$ &
$ T^{\pi}_{\rm cell}$ & $ T^{\pi}_{\rm box}$ \\
fm/$c$ &  MeV  &  MeV  &  MeV  &  MeV  &  MeV  \\
\tableline\tableline
10 & 160.6 & 121 & 125 & 99 & 106 \\
11 & 155.2 & 114 & 120 & 96 & 104 \\
12 & 150.5 & 108 & 118 & 92 & 104 \\
13 & 146.5 & 121 & 117 & 89 & 103 \\
14 & 142.4 & 113 & 112 & 86 & 102 \\
15 & 138.5 & 108 & 113 & 85 & 101 \\
\end{tabular}
\label{tab2}
\end{table}

\begin{table}
\caption{
The mean number of elastic collisions ($N_{\rm elas.}$), inelastic
reactions proceeding via the string fragmentations 
($N_{\rm inel.}^{S}$) and resonance decays ($N_{\rm inel.}^{R}$),
and via other inelastic channels ($N_{\rm inel.}^{\rm other}$) per
fm/$c$ in the UrQMD box with $V = 125$ fm$^3$ at the quasiequilibrium 
stage. Energy density, baryon density, and strangeness density in the 
box are the same as those in the central cell in central Au+Au 
(Pb+Pb) collisions at AGS (SPS) energy at $t=13$ (10) fm/$c$.
}

\begin{tabular}{c|cccc|ccccc}
 {\rm Final} & \multicolumn{4}{c} {\rm AGS} ($t=13$ fm/$c$) & 
  \multicolumn{4}{c} {\rm SPS} ($t=10$ fm/$c$) \\
 {\rm mult.} & $N_{\rm elas.}$ & $N_{\rm inel.}^{S}$ &
 $N_{\rm inel.}^{R}$ & $N_{\rm inel.}^{\rm other}$ & 
 $N_{\rm elas.}$ & $N_{\rm inel.}^{S}$ &
 $N_{\rm inel.}^{R}$ & $N_{\rm inel.}^{\rm other}$ \\
\tableline\tableline
 1 &     &     &     &36.94&      &     &      &170.26 \\
 2 &30.14& 2.26&33.05&11.91&151.83&17.87&130.61& 3.773 \\
 3 &     & 2.32& 0.65&     &      &32.57& 1.907&       \\
 4 &     &0.067&0.037&     &      & 0.81& 0.520&       \\
 5 &     &0.004&     &     &      & 0.04&      &       \\
 6 &     &0.002&     &     &      & 0.01&      &       \\
\end{tabular}
\label{tab3}
\end{table}

\end{document}